\documentclass[12pt]{elsart3p}
\usepackage{graphicx,amssymb}
\usepackage{color}
\def\bq{\begin{equation}}
\def\eq{\end{equation}}
\def\bqy{\begin{eqnarray}}
\def\eqy{\end{eqnarray}}
\def\bqyn{\begin{eqnarray*}}
\def\eqyn{\end{eqnarray*}}
\def\bi{\begin{itemize}}
\def\ei{\end{itemize}}
\def\bc{\begin{center}}
\def\ec{\end{center}}

\def\nms{\mathsurround=0pt}
\def\overapprox#1#2{\lower 2pt\vbox{\baselineskip 0pt \lineskip - 1pt
     \ialign{$\nms#1\hfil##\hfil$\crcr#2\crcr\approx\crcr}}}
\def\oversim#1#2{\lower 2pt\vbox{\baselineskip 0pt \lineskip 1pt
     \ialign{$\nms#1\hfil##\hfil$\crcr#2\crcr\sim\crcr}}}

\begin{document}
\begin{frontmatter}
\title{\large\bf Away-side distribution in a parton multiple scattering model and background-suppressed measures}
\author[ut] {\normalsize Charles B. Chiu} and
\author[uo] {\normalsize  Rudolph C. Hwa}

\address[ut]{ Center for Particle Physics and Dept. of physics, University of Texas at Austin, Austin, Texas 78712-0264, USA}
\address[uo]{Institute of Theoretical Science and Dept. of physics, University of Oregon, Eugene, OR 97403-5203, USA}
\begin{abstract}
A model of parton multiple scattering in a dense and expanding medium is described. The simulated results reproduce the general features of the data. In particular, in the intermediate trigger momentum region there is a dip-bump structure, while at higher trigger momentum the double bumps merge into a central peak. Also, a new measure is proposed to quantify the azimuthal distribution with the virtue that it suppresses the statistical fluctuations event-by-event, while enhancing the even-structure of the signal. 
\end{abstract}
\end{frontmatter}

\section{Introduction}
The study of correlations of jets in heavy-ion collisions has been a very active area of research for a number of reasons. From the theoretical perspective back-to-back jets are the most effective means to probe a dense medium by jet quenching \cite{xnw}.  In experiments the measurement of the azimuthal distribution in events triggered by high-$p_T$ particles has yielded bountiful data that provide a rich source of information on correlations \cite{xx}, the properties of which are largely not satisfactorily explained by QCD. It is in that subject where we present two pieces of work that are related, but independent.

First, we present a model of multiple scattering of a parton as it traverses a dense and expanding medium. An event generator is constructed to reproduce the $\Delta\phi$ distribution at intermediate trigger momentum so that the recoil partons can give rise to deflected jets in addition to the possibility of being totally absorbed. It is then extended to higher trigger momentum to show that the results are in accord with the data that exhibit punch-through jets. 

The second part of our presentation addresses the issue of how to improve the measure of the event structure by eliminating the necessity of making background subtraction, a procedure that is thus far always used in the analyses of data. We make use of factorial moments to suppress the statistical fluctuations. The proposed method is applied to the analysis of the away-side $\Delta\phi$ distribution with notable success on simulated events.

\section{A multiple scattering model}

In heavy-ion collisions the events with trigger momentum in the $2.5<p_T<4$ GeV/c range and the away-side azimuthal distribution shown in Fig. 1 \cite{phenix06} have stimulated a great deal of attention. There is a dip  at $\varphi =\Delta\phi-\pi=0$ and  double bumps at $\varphi\sim \pm 1$ rad.  This dip-bump structure has been suggested as a signature of collective response of the medium in the form of a Mach Cone structure, initiated by sonic boom, Cherenkov radiation of gluons, or other mechanisms 
\cite{casalderrey-solana06,koch05,dremin06,shuryak06}. 
Events with simultaneous presence of two associated particles approximately symmetric about $\varphi=0$ have been observed \cite{ulery06}, which support the existence of Mach cone type of events. On the other hand, we find it to be plausible that there can be a substantial fraction of events where the dip-bump structure is due to the effect of multiple scattering \cite{chiu06}.

\begin{figure}
\begin{center}
\includegraphics*[width=8cm]{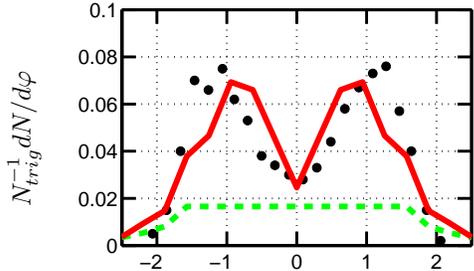}
\end{center}

\vspace{-10mm}
\caption{Dip-bump structure in the away side distribution. The associated particle angle $\varphi$ is defined with respect to the direction opposite to the trigger. The data points (solid dots) are from PHENIX collaboration\cite{phenix06} measured in Au+Au collision at 0-5\% centrality.
Momentum ranges for the triggers are 2.5-4 GeV/c and for the associated particles are 1-2.5 GeV/c.}
\label{fig:fig1}
\end{figure}

\vspace{1mm}
We assume that each triggered event of a central collision is  the result of a hard scattering near the surface on the trigger side, randomly initiated within the region 
\bq
\xi R_0<r<R_0,\,\,{\rm and}\,\,|\Delta\phi|<\phi_c.
\eq
where $R_0$ is the radius of the colliding nuclei, $\xi$ and $\xi_1=\sin\phi_c$ are close to 1.
The recoil parton  is directed opposite to the trigger. The hard collision occurs at i=1 followed by scatterings at i=2, 3, $\cdots$.  The medium undergoes Hubble-like expansion with the transverse radius \cite{koch03}
\bq
R_i=R_0exp[H/(\tau_i-1)],
\eq
and the density is assumed to take on the form
\bq
\rho_i(r_i,\tau_i)=(\rho_s/\tau_i)\theta(R_i-r_i).
\eq
The central characteristic of our model is that the recoil parton undergoes multiple scattering with
random scattering angle chosen from the forward cone defined by a Gaussian distribution with a width 
\bq
\sigma_i=\sigma_s\frac{\rho_i/\rho_s}{E_i/E_s}
\eq
We adopt the BDMPS\cite{baier01} form for the energy loss
\bq
dE/dx=\kappa_1\sqrt{E/E_s}.
\eq
The stepsize depends on the instantaneous density $\rho_i$ and parton energy $E_i$. We assume the form
\bq
\Delta_i=\Delta_s \exp(\rho_s-\rho_i)\sqrt{E_i/E_s},
\eq
where $\rho_s$ sets the dimensionless scale of the density dependence.
Eqs(5) and (6) together lead to the recursion relation for the parton energy
\bq
E_{i+1}=E_i[1-\kappa \exp(\rho_s-\rho_i)]^2,
\eq
with $\kappa=\kappa_1\Delta_s/2E_s$. The  time interval between scattering is
\bq
\tau_{i+1}-\tau_i=\Delta_i/\Delta_1.
\eq
If the parton energy is below the cutoff $E_{\rm cut}$, the parton is absorbed by the medium. 

In our simulation, the effective trigger momentum is assumed to be at 4.5GeV and the exit parton momentum range 0.7 to 2.2 GeV/c for $1<p_T({\rm assoc})<2.5$ GeV/c.  The shift of 0.3 GeV/c is to account for the recombination of exit parton with at least one thermal parton which on average carries a momentum of 0.3 GeV/c.  

Following parameters are fixed at their default values:
$R_0=6$f, $\xi=\xi_1=0.8$. H=0.07 (see ref.\cite{koch03}), $E_s$=5 GeV. $E_{\rm cut}$=0.3 GeV/c.  Other parameters are adjusted so that the exit track distribution together with a thermal background qualitatively reproduce the dip-bump feature of the data. The four adjustable parameters are: $\rho_s=0.63$, $\sigma_1= 0.88$ rad, $\Delta_s=1.9$f and $\kappa=0.17$. 
The dashed line in Fig.\ 1 is the thermal background due to the energy deposited in the medium \cite{chiu06}.  
The solid line rises above the dashed line due to the contribution from the $\varphi$ distribution of exit partons multiplied by a factor of 1.7 to account for the average number of hadrons produced from each exit parton through recombination of possibly more than one shower parton with thermal partons.
 We regard the agreement with data in Fig. 1 as satisfactory.

Sample tracks are shown in Fig 2. Both plots in the figure are superpositions of many events, with one track per event. Exit tracks correspond to those persistently bending away from the center. These tracks are relatively short, and leave the medium at $\varphi\approx 1$.  The absorbed tracks tend to swing back and forth,   and are relatively long.  It is this underlying pattern that leads to the dip-bump structure at the intermediate energy.  
\begin{figure}
\begin{center}
\includegraphics*[width=7cm]{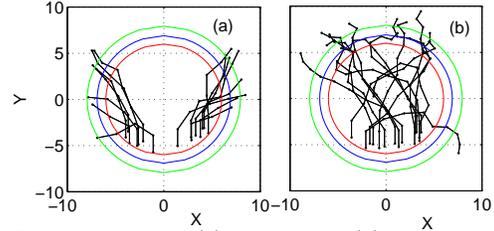}
\end{center}

\vspace{-6mm}
\caption{Samlple tracks. (a) Exit tracks. (b) Absorbed tracks.}
\label{fig:fig2}
\end{figure}

 At higher trigger momentum the data  reveal  only one central peak as shown in  Fig.~3.  Since the exit partons take more than 10 fm/c to traverse the medium, the thermal partons on the far side are significantly reduced at such later times. Thus the  hadronization of the exit partons is dominated by fragmentation, when the associated hadron momentum is above 4 GeV/c. The predicted hadron distributions are shown by the solid curves in Fig.~3.

Our  results indicate that the present multiple scattering model can provide a  good description of the main features of the data, i.e. the dip-bump structure at lower trigger momentum, and the merging of double bumps into one central peak at high trigger momentum. The important consequence is that only one jet on the away side, if any, accompanies a near-side jet, not two.
\begin{figure}
\begin{center}
\includegraphics*[width=7cm]{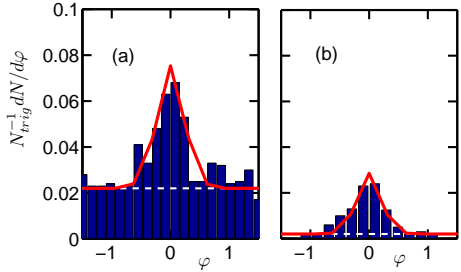} 
\end{center}
\caption{Comparison between the model fits to the peaks and the data. The data points are from STAR collaboration\cite{star06} in momentum ranges of the triggers: 8-15 GeV/c and of the associated particles for (a) 4-6 GeV/c and for (b) $>6$ GeV/c.}
\label{fig:fig3}
\end{figure}

\section{Use of the background suppressed measures to analyze away-side distribution}
So far we have focused on only the event-averaged data. Next we investigate the implication of the event-by-event description of the model. We will use the factorial moment (FM) approach to analyze the data. 
Consider an event with multiplicity $N$ on the away side. We divide the $\varphi$-region of interest into $M$ bins. 
FM of order $q$ is defined by
\bq
f_q=\frac{1}{M} \Sigma_j n_j(n_j-1)\dots (n_j-q+1),\,\, n_j\ge q.
\eq
Normalized factorial moment (NFM) is defined by
\bq
F_q=f_q/f_1^q.
\eq

The statistical-limit theorem states that if in each bin the particle number  has a Poisson-like fluctuation in the large $N$  limit, then $F_q$ approaches 1 for all $q$, independent of $M$.  The NFM is defined for each event, and involves an average over bins. 
In practice $N$  is finite and $F_q$ fluctuates from event to event. Fig 4(a) displays the fluctuation of $F_3$ for some 500 background events. The distribution of $F_3$, i.e. $dN/dF_3$ vs $F_3$, is shown in Fig.4 (b) as the solid curve. The dashed curve for $F_2$ is sharper and peaks closer to the ideal limit of unit. We will see that it is the event-averaged  NFM that constitutes the basic background suppressed measure \cite{chiu06a}. 
 \begin{figure}
\begin{center}
\includegraphics*[width=7cm]{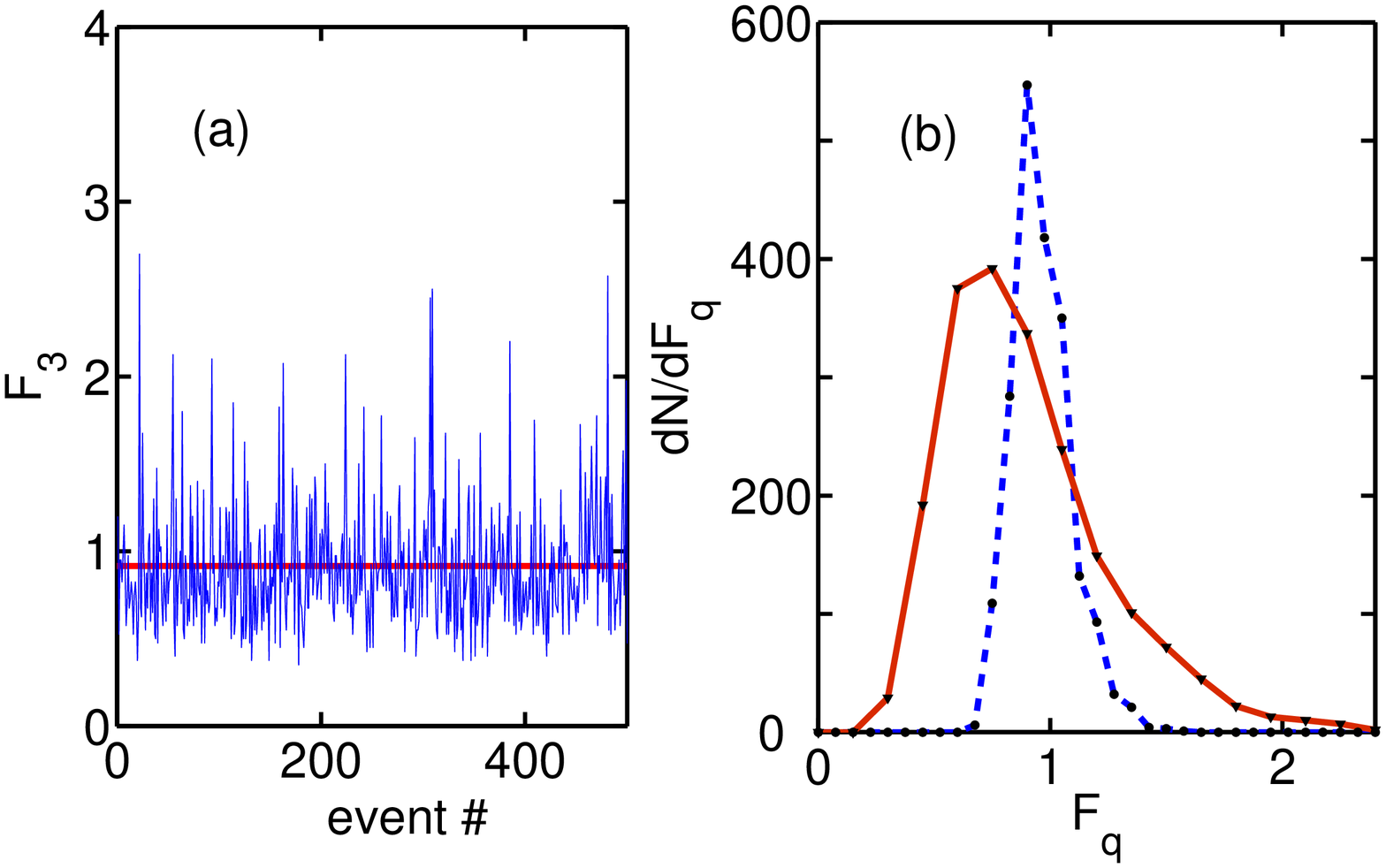}
\end{center}
\caption{(a) Illustration of event-by-event fluctuations of $F_3$. (b) A comparison of dispersion curves between $F_2$, and $F_3$. The $F_2$ curve is sharper with its peak close to the statistical limit of unity.} 
\label{fig:fig4}
\end{figure}

We now turn to a toy model to illustrate the uses of the FM method. Here the signal is defined by a cluster of several particles spread over a small $\varphi$ interval. We will refer it as a jet. For our model calculation we use several particles evenly spaced in a small $\varphi$ interval. Three type of events are considered. First, the background(bg) events. Each bg-event consists of large number of particles randomly distributed over the full $\varphi$ interval of interest. The second is background plus one jet (bg+1j) events. The jet location is uniformly distributed in a subdomain within the full range. They mimic deflected jet plus background events. The third type is (bg+2j) events, where the 2 jets are symmetrically produced about $\varphi=0$ at small $\varphi$ region,  representing the Mach-cone type events. A typical superposed spectra  of all the three types of events are shown in Fig.~5.
\begin{figure}
\begin{center}
\includegraphics*[width=5cm]{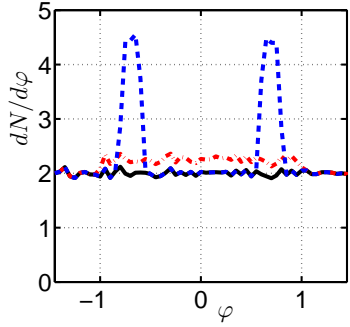}
\end{center}

\vspace{-3mm}
\caption{A toy model: Superpositons of $\varphi$ distributions for the production of bg+2j(dash), bg+1j(dash-dot) and bg (solid-line) events.} 
\label{fig:fig5}
\end{figure}  

Fig. 6 gives the $\left<F_q\right>$ vs $M$ plots. Plot (a) shows the average  NFM of the background events. Here we have $\left<F_q\right>\sim1$ for the range of $M$ shown. While $F_q$  varies from event to event,  the event-averaged quantity $\left<F_q\right>$ is approximately the statistical limit.    Sensitivity of $\left<F_q\right>$ to signals can be seen in plots (b) and (c). Deviations of $\left<F_q\right>$ from unity become noticeable for $q$=3 and 4,  especially at high $M$.
\begin{figure}
\begin{center}
\includegraphics*[width=7.5cm]{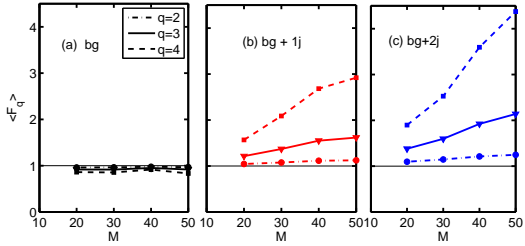}
\end{center}
\caption{The $M$-dependence of $\left<F_q\right>$ for various orders of $q$. (a) bg,  (b) bg+1j, and (c) bg+2j.} 
\label{fig:fig6}
\end{figure}  

Factorial moment approach can also be used to measure the fluctuations between two $\varphi$-regions. Consider the case where region-I is for $\varphi<0$, and region-II, for $\varphi>0$. Here the difference measures the fluctuation. Define \bq
\left<D(p,q)\right>=\left <|F_q^I-F_q^{II}|^p\right>.
\eq
 where the $p$th power amplifies the fluctuations. To track the relative normalization one also needs to define the corresponding  sums
\bq
\left<S(p,q) \right>= \left<(F_q^I+F_q^{II})^p\right>.
\eq
The $\left<D\right>$  vs $\left<S\right>$ plots are shown in Fig.~7. A glance at the figure shows that there is a common pattern in all plots. More specifically in each plot the bg contribution is well localized and suppressed. The bg+1j and bg+2j contributions fan out along approximately straight lines with bg+1j events having larger slopes than those  of the bg+2j events. Notice that the coordinate scales increase quickly as $p$ and $q$ increase. As mentioned earlier, the bg+2j case mimics the Mach cone type of events, while the bg+1j case the deflected jet events. One sees that  the $\left<D\right>$  vs $\left<S\right>$  plot can be used to distinguish between the two types of events without making background subtractions. 
 \begin{figure}
\begin{center}
\includegraphics*[width=8cm]{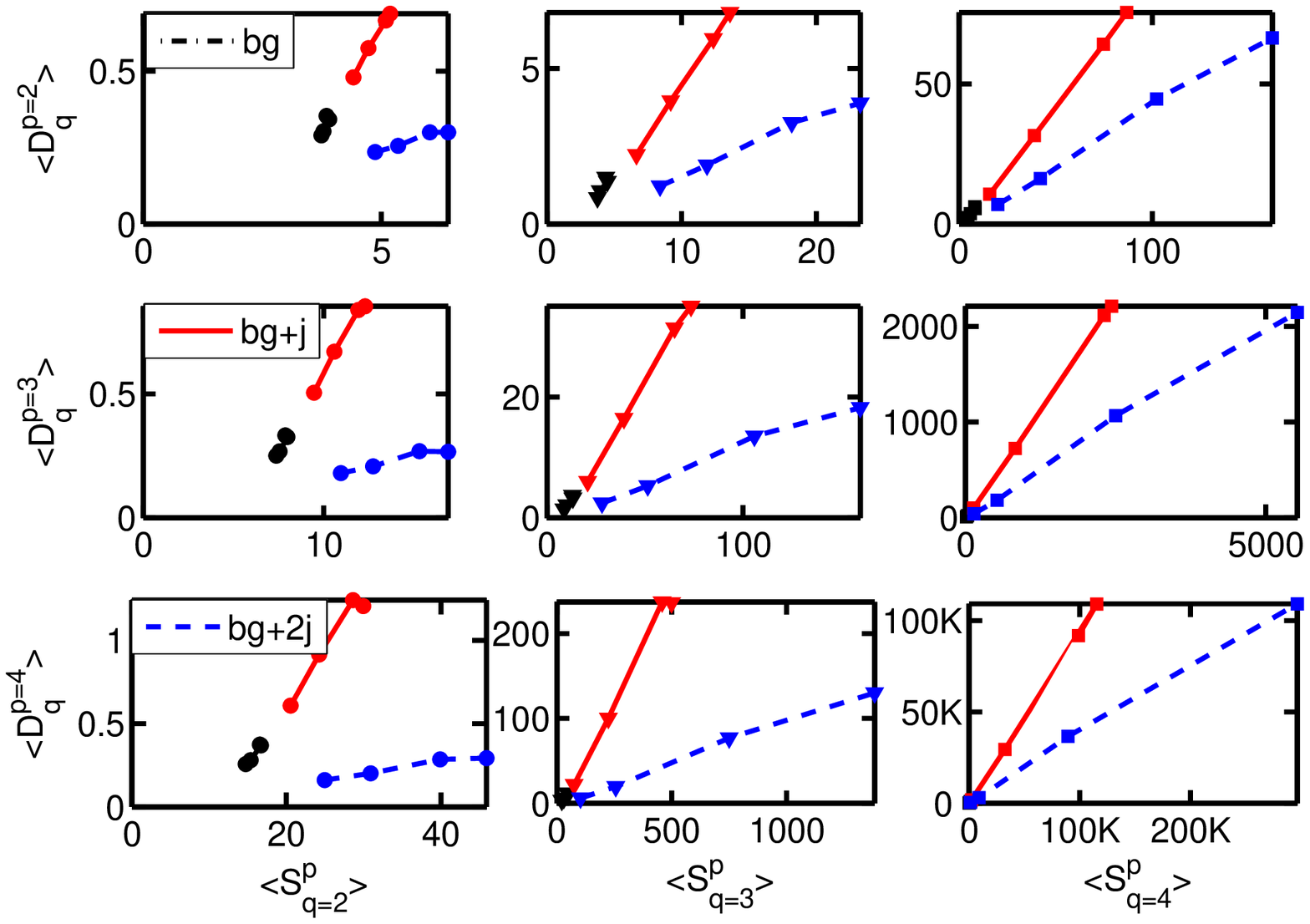}
\end{center}
\caption{The $\left<D(p,q) \right>$ vs $\left<S(p,q) \right>$ plots. Rows: $p$=2,3 and 4, columns: $q$=2, 3 and 4. }
\label{fig:fig7}
\end{figure}  

To summarize, we have investigated the use of FM method to analyze away-side $\Delta\phi$ distribution. The main advantage is that in plots involving FM measures signals tend to stand out and the background contribution is suppressed. And there is no need to make background subtraction. We envision that the FM-method has the potential to provide a common framework to compare results from different experiments and various subtraction schemes.

\end{document}